# Biophysical interaction, nanotoxicology evaluation, and biocompatibility and biosafety of metal nanoparticles

**Catalano Enrico**[1]

[1] Italian Institute of Technology, Via Morego, 30, 16163 Genova, Italy

**Corresponding author**

Catalano Enrico

Italian Institute of Technology, Via Morego, 30, 16163 Genova, Italy. email: enrico.catalano@medisin.uio.no

**Abstract**

Nanotechnology has been one of the fastest growing fields in the last three decades. Nanomaterials (sized 1-100 nm) has a wide spectrum of potential applications in many fields, applied as coating materials or in treatment and diagnosis. Nowadays, nanoparticles of both metallic and non-metallic origin are under investigation and development for applications in various fields of biology/therapeutics. Specifically, we show the correlations between the physicochemistry and biophysical specificity of metal nanoparticles and their uptake, transport, and biodistribution in cells, at the molecular, cellular, and whole organism level.

Physiologically important metals are present in the human body with a wide range of biological activities. Some of these metals are magnesium, chromium, manganese, iron, cobalt, copper, zinc, selenium and molybdenum. Metals used in nanotechnology have to be biocompatible with the human system in terms of absorption, assimilation, excretion, and side effects.

These metals are synthesized in the form of nanoparticles by different physical and chemical methods. Nanotoxicological studies of metal nanoparticles are intended to determine whether and to what extent their properties may pose a threat to the environment and to human beings. An overview of metal and metal oxide nanoparticles, their applications, and the potential for human exposure is provided, and it is integrated by a discussion of general principles of nanoparticle-induced toxicity and methods for toxicity testing of nanomaterials. This review wants to focus on establishing metal nanoparticles of physiological importance to be the best candidates for future nanotechnological tools and medicines, owing to the acceptability and safety in the human body. This can only be successful if these particles are synthesized with a better biocompatibility and low or no toxicity.

# 1. Nanotechnology: an overview

Nanotechnology is an evolving scientific field that has allowed the manufacturing of materials with novel physicochemical and biological properties, offering a wide spectrum of potential applications. Properties of nanoparticles that contribute to their usefulness include their markedly increased surface area in relation to mass, surface reactivity and insolubility, ability to agglomerate or change size in different media and enhanced endurance over conventional-scale substance. Nanoparticle applications are in several fields; from active food packaging to drug delivery and cancer research.

Recent applications of nanoscience include the use of nanoscale materials in electronics, catalysis, and biomedical research. Among these applications, strong interest has been shown to biological processes such as blood coagulation control and multimodal bioimaging, which has brought about a new and exciting research field called nanobiotechnology.

Nanotechnology has exciting therapeutic applications, including novel drug delivery for the treatment of cancer. This growth in nanomedicine also fuels advances in bioengineering, regenerative medicine and the development of medical devices.

The very dynamic growth of nanomedicine and their medical applications over the past 15 years has promised to add a new impetus to the diagnostics and therapeutics of a wide range of human pathologies, including cancer, cardiovascular diseases and diseases of the central nervous system. Properties of nanoparticles that are critical to their utilization include their surface reactivity and insolubility, markedly increased surface area in relation to mass, ability to agglomerate or change size in different media and enhanced endurance over conventional-scale substance. The rapid production and incorporation of engineered nanomaterials into consumer products alongside research suggesting detailed characterization of nanomaterials with in vitro assays desirable for nanosafety screening. There is a growing concern on potential toxicity and adverse effects of nanomaterials on human health, including lack of standard method of assessment of toxicology of these materials. Multifunctional nanoparticles including metal composite nanomaterials and ferromagnetic cores rely on their ability to interact efficiently with electromagnetic (EM) fields to produce a response through the scattering or absorption of the interacting field [1]. EM-mediated responses may be used for selective detecting, targeting, monitoring and treating a wide cross section of human diseases. However, the impact of an electromagnetic field on interactions between multifunctional nanoparticles, cells and subcellular structures needs to be carefully evaluated to determine the risk/benefit ratio of such pharmaceuticals.

## 1.1 Biophysical interaction of nanoparticles

A crucial increasing effort is focused on biophysical interaction of nanoparticles at large. When a NP docks on cell membrane, it forms a highly heterogeneous NP-cell interface and initiates dynamic physiochemical interactions and a sequence of kinetic processes.

The biophysical interaction of different origins shape the interactions, modify the associated energy landscapes, and dictate the endocytosis of the NPs. Wrapping NPs necessitates involving the cell membrane and pulling the membrane against membrane tension to the wrapping site. All the other forces, including electrostatic, van der Waals (vdW), hydrophobic forces, ligand-receptor binding etc., are the key-elements for adhesion force that drives endocytosis. The adhesion force can stem from either specific or nonspecific interactions, or both. Specific interactions involve recognition and binding of the ligands coated on the NP surface to the complementary receptors on the cell membrane. The applications of nanomaterials are indeed aplenty: tear resistant carbon nanotube fabrics to carbon nanotube transistors, displays and radios, to a contrast enhancing agent for MRI, to graphene vapor sensors, quantum dot tags, nanopharmaceuticals and dendritic drug and prodrug transporters, the potential of nanomaterials expands rapidly over time.

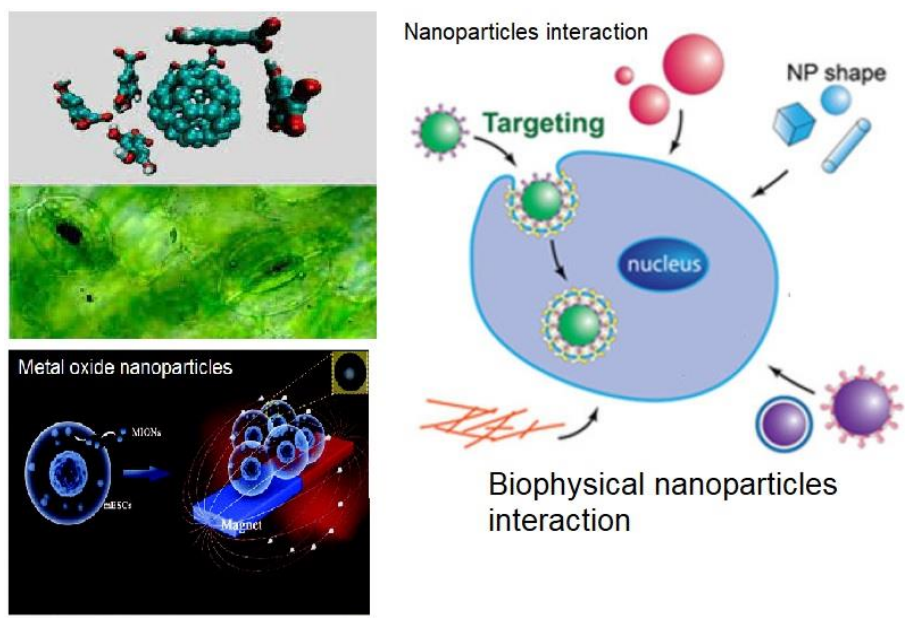

## 2. Nanotoxicology of metal nanoparticles

Nanotoxicology is a sub-specialty of particle toxicology, which investigates the toxicity of nanoparticles and nanomaterials. Nanomaterials have unique properties compared with their larger counterparts related to quantum size effects and large surface area to volume ratio. Metallic nanoparticles are used for scientific and medical purposes. Typical metal nanoparticles that have been studied are titanium dioxide, alumina, zinc oxide, iron-oxide, manganese, etc. Metallic nanoparticles have much larger surface area to unit mass ratios which in some cases may lead to greater pro-inflammatory effects. Many metals show a strange dose-response pattern comprising stimulation of immune function at low doses and suppression at higher doses, but overall immune function is often preserved due to the redundancy and the reserve capacity of the immune system, and clinically relevant effects do not manifest themselves. In addition, some nanoparticles seem to be able to translocate from their site of deposition to distant sites such as the blood and the brain. This has resulted in a sea-change in how particle toxicology is viewed- instead of being confined to the lungs, nanoparticle toxicologists study the brain, blood, liver, skin and gut.

The toxicology of nanoparticles (particles <100 nm diameter) appears to have toxicity effects that are unusual and not seen with larger particles and particle size is likely to contribute to cytotoxicity. Many factors such as size, inherent properties, and surface chemistry may cause nanoparticle toxicity. Conventional physical and chemical methods of metal nanoparticle synthesis may be one possible reason for nanoparticle toxicity that can be overcome by synthesis of nanoparticles from biological sources. The physicochemical properties of nanoparticles establish how they interact with cells and can induce their overall potential toxicity. Recent studies have started investigating various characteristics that make some nanoparticles more toxic than others. Understanding these mechanisms and details can lead to the development of safer nanoparticles.

New opportunities are emerging from the development of metallic multifunctional nanomaterials and nanoparticles for diagnostics and therapy, and it is important to investigate all related problems of nanotoxicity with the aim to have biocompatible nanomaterials for future applications of nanomedicine devices. Nanomaterials, even when made of inert elements such as gold, become highly active at nanometer dimensions. Nanotoxicological studies are used to determine whether and to what extent these properties may pose a threat to the environment and to human health [2]. The combination of different metals with each other in nanomaterials can cause complex toxicity, which does not occur with single metals. In this regard the effect of oxidative stress caused by asbestos is known as the main factor in asbestosis and in disturbing cell structure and additionally the effects of nanoparticles on respiratory toxicity and inflammation [3]. Different engineered nanomaterials are present in everyday life used in products with direct exposure to humans. Several examples can be illustrated on this: $Fe_2O_3$ nanoparticles that are used in the final polish on metallic jewelry, $TiO_2$ nanoparticles are used in food coloring, cosmetics, skin care products, and tattoo pigment [4-6], ZnO nanoparticles present in many products including cotton fabric, food packaging, and rubber for its deodorizing and antibacterial properties [7, 8].

**2.1 Immunotoxicology of Metals**

Immune reaction plays a remarkable role in toxicity, which is important for toxicologists, especially in respiratory diseases. Clinically relevant hypersensitivity reactions due to metals are dominated by T cell-mediated allergic contact dermatitis, particularly in response to exposure to beryllium, cobalt, chromium, gold, mercury, and nickel. Immediate (type I) hypersensitivity reactions dominated by airways symptoms occur infrequently, and then most often with platinum, but rarely with nickel or chromium. The induction of metal-induced autoimmunity, including the formation of immune-complex deposits, is well documented in humans, but the number of recognized cases is few.

The size of nanoparticles is an important factor in the immune system activation. Some studies on the different sizes of carbon and titanium oxide showed that reduction in nanoparticle size increases its toxicity in the lungs. Some of the particle features such as size, surface chemistry, and oxidative stress functions play important roles in nanotoxicity. Other features such as crystallinity, coating, and the longevity of particles have also been studied as important parameters [9]. By gaining control over dangerous particles, the aim is the use of nanoparticles by reducing their harmful effects, and thus allowing them to be used in the curing of diseases [9-12].

**3. Important factors involved in metallic nanotoxicology**

Nanotoxicology, the study of nanomaterial toxicity to biological systems, has become a vibrant area of research over the past 10 years. Two main factors drive this investigation: (1) industrial interest in engineered nanomaterials [13], leading to their increased use in consumer products and increased possibility of human and environmental exposure, and (2) the unique physical and chemical properties of nanomaterials relative to their bulk counterparts [14]. The effects of metallic NPs exposure may involve toxicity induced by multiple mechanisms and it can be classified into primary and secondary categories (Figure 1), depending upon the extent of exposure.

The main metals in the form of nanoparticles and engineered nanomaterials used in products with direct exposure to humans, that mediate toxicity, belongs to the fourth period of the periodic table of elements [76]. They are Ti, Cr, Mn, Fe, Ni, Cu, and Zn and their related seven oxides of transition metals (Table 1).

Primary effects derived from direct cellular NP contact may include toxicity, oxidative stress, altered signaling pathways that perturb cellular homeostasis leading to cellular injuries, DNA damage, and inflammation [77-79]. This can lead to suppression of proliferation (via cell cycle arrest). When cells cannot overcome the stress and fix the damage, they are destined to death (apoptosis or necrosis). Due to their small size, metallic NPs may translocate through tissue barriers into the blood, where they can circulate and eventually deposit in other organs, thereby generating a secondary NP exposure.

Secondary effects are connected to toxicity at the site of metallic NPs deposition, in organs such as the liver, spleen, or kidneys, inducing stimulation of systemic inflammation or alterations in systemic function [77-79]. The routes of exposure for human body derive from external metallic NP sources, and also from internal exposure, for instance when orthopedic or surgical implant wear metallic NPs are released locally from the implant site [80-82]. In most of the cases, humans are exposed to metallic NPs in their environments through the pulmonary route, by inhaling airborne NPs during normal breathing [83].

| Metals | Industrial Applications and Uses |
|---|---|
| **Titanium (Ti)** | White pigment, white food coloring, cosmetic and skin care products, thickener, tattoo pigment and styptic pencils, plastics, semiconductor, solar energy conversion, solar cells, solid electrolytes, detoxification or remediation of wastewater; used in resistance-type lambda probes; can be used to cleave protein that contains the amino acid proline at the site where proline is present |
| **Chromium (Cr)** | Protection of silicon surface morphology during deep ion coupled plasma etching of silica layers; used in paints, inks, and is the precursor to the magnetic pigment chromium dioxide. Magnetic tape emulsion, data tape applications. |
| **Manganese (Mn)** | Electrochemical capacitor, as a catalyst; used in industrial water treatment plants |
| **Iron (Fe)** | Used as contrast agents in magnetic resonance imaging, in labeling of cancerous tissues, magnetically controlled transport of pharmaceuticals, localized thermotherapy, preparation of ferrofluids, final polish on metallic jewelry and lenses, as a cosmetic, tattoo inks.<br>MRI scanning, and as a black pigment. |
| **Nickel (Ni)** | In ceramic structures, materials for temperature or gas sensors, nanowires and nanofibers, active optical filters, counter electrodes Electrolyte in nickel plating solutions; an oxygen donor in auto emission catalysts; forms nickel molybdate, anodizing aluminum, conductive nickel zinc ferrites; in glass frit for porcelain enamel; thermistors, varistors, cermets, and resistance heating element. |
| **Copper (Cu)** | Burning rate catalyst, superconducting materials, thermoelectric materials, catalysts, sensing materials, glass, ceramics, ceramic resisters, magnetic storage media, gas sensors, near infrared tilters, photoconductive applications, photothermal applications, semiconductors, solar energy transformation; can be used to safely dispose of hazardous materials Pigment, fungicide, antifouling agent for marine paints, semiconductor |
| **Zinc (Zn)** | Added to cotton fabric, rubber, food packaging, cigarettes, field emitters, nanorod sensors; Applications in laser diodes and light emitting diodes (LEDs), a biomimic membrane to immobilize and modify biomolecules; increased mechanical stress of textile fibers. |

88

| Elements | Oxide | Potential Application |
|---|---|---|
| Scandium (Sc) | $Sc_2O_3$ | Used in high-temperature systems for its resistance to heat and thermal shock, electronic ceramics, and glass composition |
| Titanium (Ti) [1–7] | $TiO_2$ | White pigment, white food coloring, cosmetic and skin care products, thickener, tattoo pigment and styptic pencils, plastics, semiconductor, solar energy conversion, solar cells, solid electrolytes, detoxification or remediation of wastewater; used in resistance-type lambda probes; can be used to cleave protein that contains the amino acid proline at the site where proline is present, and as a material in the meristor |
| Vanadium (V) | $V_2O_5$ | Catalyst, a detector material in bolometers and microbolometer arrays for thermal imaging, and in the manufacture of sulfuric acid, vanadium redox batteries; preparation of bismuth vanadate ceramics for use in solid oxide fuel cells [8] |
| | $V_2O_3$ | Corundum structure as an abrasive [9], antiferromagnetic with a critical temperature at 160 K [10] can change in conductivity from metallic to insulating |
| Chromium (Cr) | $Cr_2O_3$ | Protection of silicon surface morphology during deep ion coupled plasma etching of silica layers; used in paints, inks, and is the precursor to the magnetic pigment chromium dioxide |
| | $CrO_2$ | Magnetic tape emulsion, data tape applications |
| Manganese (Mn) | $MnO_2$ | Electrochemical capacitor, as a catalyst; used in industrial water treatment plants |
| Iron (Fe) | $Fe_2O_3$ | Used as contrast agents in magnetic resonance imaging, in labeling of cancerous tissues, magnetically controlled transport of pharmaceuticals, localized thermotherapy, preparation of ferrofluids [11,12], final polish on metallic jewelry and lenses, as a cosmetic |
| | FeO | Tattoo inks |
| | $Fe_3O_4$ | MRI scanning [13], as a catalyst in the Haber process and in the water gas shift reaction [14], and as a black pigment [15] |
| Cobalt (Co) | $Co_2O_3$ | Catalyst; for studying the redox and electron transfer properties of biomolecules; can immobilize protein |
| | CoO | Blue colored glazes and enamels, producing cobalt(II) salts |
| Nickel (Ni) | NiO | In ceramic structures, materials for temperature or gas sensors, nanowires and nanofibers, active optical filters, counter electrodes |
| | $Ni_2O_3$ | Electrolyte in nickel plating solutions; an oxygen donor in auto emission catalysts; forms nickel molybdate, anodizing aluminum, conductive nickel zinc ferrites; in glass frit for porcelain enamel; thermistors, varistors, cermets, and resistance heating element |
| Copper (Cu) | CuO | Burning rate catalyst, superconducting materials, thermoelectric materials, catalysts, sensing materials, glass, ceramics, ceramic resisters, magnetic storage media, gas sensors, near infrared tilters, photoconductive applications, photothermal applications, semiconductors, solar energy transformation [16]; can be used to safely dispose of hazardous materials [17] |
| | $Cu_2O$ | Pigment, fungicide, antifouling agent for marine paints, semiconductor |
| Zinc (Zn) | ZnO | Added to cotton fabric, rubber, food packaging [18–20], cigarettes [21], field emitters [22], nanorod sensors; Applications in laser diodes and light emitting diodes (LEDs), a biomimic membrane to immobilize and modify biomolecules [23]; increased mechanical stress of textile fibers [24] |

**Table 1.** Metallic nanoparticles and engineered metallic nanomaterials used in products with direct exposure to humans.

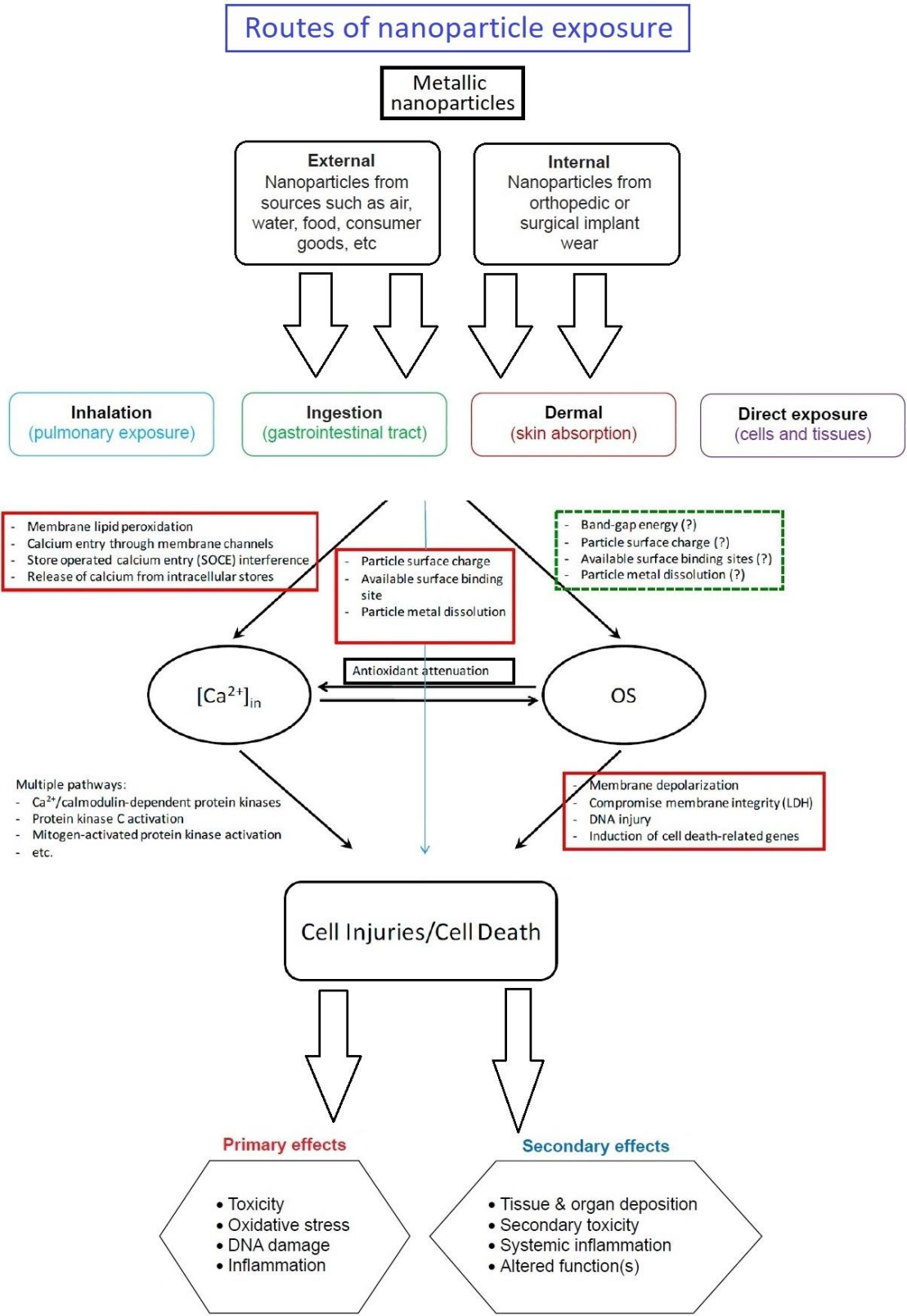

**Figure 1.** Routes of metallic nanoparticle exposure and cell toxicity effects.

## 3.1 Aggregation and degradation of nanoparticles

Transmission electron microscopy (TEM) remains the gold-standard for characterizing the primary size and structure of nanomaterials based on its high spatial resolution. Chemical information about a nanomaterial can also be obtained through parallel analysis using spectroscopic methods such as electron energy loss spectroscopy (EELS) or energy-dispersive X-ray spectroscopy (EDX). Despite this versatility, sample preparation requirements for TEM traditionally limits its utility for characterizing metallic nanomaterial dispersions in situ (i.e., under biologically or environmentally relevant conditions). In particular, the high-vacuum environment of electron microscopes can lead to sample preparation artifacts (e.g., nanomaterial aggregation induced by solvent evaporation) that do not reflect nanomaterial behavior in solution. However, advances in sample preparation are expanding the utility of electron microscopy for in situ nanomaterial characterization. Cryo-TEM, in which samples are imaged in a vitrified state, preserves the spatial arrangement of species as they exist in solution. While cryo-TEM is a powerful probe of nanoscale packing, it cannot asses the dynamic behavior of nanomaterial dispersions. Liquid-cell TEM does preserve the fluid environment around the nanomaterial.

## 3.2 Nanomaterial Surface Adsorption

Metallic nanomaterials exhibit not only dynamic colloidal and physicochemical behavior under biologically relevant conditions but also dynamic changes in surface chemistry. Components of biological and environmental matrixes, including proteins [15, 16], lipoproteins [17], and dissolved organic matter [18, 19], can adsorb onto metallic nanomaterial surfaces under realistic exposure scenarios, forming a molecular "corona". Protein coronas, being the most widely studied, have been shown to significantly alter nanomaterial colloidal behavior and interaction with biological systems relative to the pristine nanomaterial by modifying factors such as nanomaterial size, surface charge, and surface chemistry [20].

The composition of the protein corona, e.g., the glycosylation state of adsorbed proteins, can influence the extent of nanomaterial internalization by cells [21, 22]. Adsorbed proteins can also influence immune response in cells, a subject that has been recently reviewed [23]. Adsorption of other molecules has also been shown to alter nanomaterial colloidal behavior and toxicity [24-26]. Detailed characterization of the dynamic composition of the protein corona, as well as coronas formed from less-studied molecules like dissolved organic matter, is necessary to understand the interface between nanomaterials and biological systems under biologically and environmentally relevant exposure scenarios. The binding constants controlling molecular adsorption to nanomaterial surfaces and the composition, surface-coverage, thickness, and orientation of molecules on the nanomaterial surface have been studied using several analytical techniques. A significant barrier to characterization of molecular adsorption to nanomaterial surfaces in situ is signal interference due to the high chemical complexity of biologically relevant media (which may contain proteins, lipids, and dissolved organic matter). Additionally, discriminating between species adsorbed to the nanoparticle surface and those in solution can be difficult due to their similar chemical signatures. These problems can be ameliorated by isolating subpopulations of metallic nanomaterials or by isolating nanomaterials from interfering colloidal species in the sample matrix prior to analysis; however, separation methods have been infrequently employed in published studies. Currently available separation methods include column chromatography and field flow fractionation and its derivatives. Development of more rapidly implementable and generalizable methods to separate nanomaterials from complex biological matrixes would facilitate accurate nanomaterial characterization.

**3.3 Nanomaterial-Induced Changes in Physiological Processes**

The full understanding of nanomaterial toxicity requires that the impacts of nanomaterials on the biomolecular components of cells must be considered in parallel with their system-wide impacts on physiological function. Perturbations to processes such as DNA replication, protein biosynthesis, and exocytosis can have significant consequences for cell or organism health and should be understood to ensure the responsible use of engineered metallic nanomaterials. Understanding nanomaterial effects on physiological processes can also provide further insight into the mechanisms underlying nanomaterial-induced toxicity. Analytical and experimental methods adopted from areas within the chemical and biological sciences are enabling investigations of nanomaterial perturbations to physiological processes across a range of organismal complexity, from single-celled organisms like bacteria to primary human cells lines and multicellular organisms like water fleas. Linking nanomaterial impacts on physiological function to underlying mechanisms of toxicity remains a major challenge for the nanotoxicity research community.

## 3.4 The clearance of nanomaterials in the human body

Metallic nanoparticles can enter in the human body via external or internal exposure, the processes involved in this intrusion are: absorption, distribution, metabolism and excretion. These processes constitute the subject of toxicokinetics (**Figure 2**). Metallic nanomaterials may enter the organism via medical intervention or dermal, oral as well as respiratory exposure. Therefore, toxicokinetics deals with the potential toxic effects of substances on the exposed tissues/organs. To date, the majority of toxicokinetic studies deal with metals or metal oxide nanomaterials and to a lesser degree with polymer nanomaterials [27, 28].

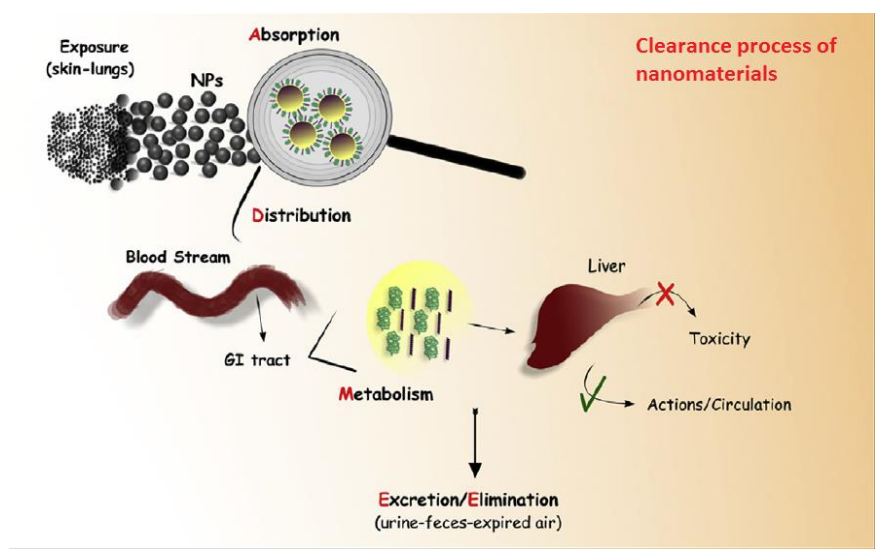

**Figure 2.** Schematic presentation of the clearance process of NPs. When NPs enter the human body, through various exposure ways.

## 4. Role of size of nanoparticles in nanotoxicology

The two main aspects which can induce nanotoxicity of nanoparticles are: size and chemical compounds. A reduction in the size of nanoparticles results in an increase in particle surface area. Therefore, more chemical molecules may attach to this surface, which would enhance its reactivity and result in an increase in its toxic effects [11, 12]. The suppression of cell proliferation induced by metallic nanoparticles occurs when cells become arrested in one or more cell cycle phases. Many studies on the absorption of metallic nanoparticles from different biological barriers have examined these effects. After absorption, nanoparticles reach the blood stream and then spread through the tissue. In one study, 33% of 50 nm, 26% of 100 nm, and 10% of 500 nm particles were discovered in mucosal and lymphatic tissues of the intestine [11]. Nanoparticles larger than 1 μm were weakly observed and nanoparticles larger than 3 μm were occasionally seen in lymphatic tissues. Researchers have concluded that:

- Metallic nanoparticles smaller than 100 nm are absorbed by the cells of the intestine but not the larger nanoparticles (300 nm).
- The absorption of smaller nanoparticles (100 nm) in the lymphatic tissue is greater than in intestinal cells.
- Intestinal cells cannot absorb metallic nanoparticles larger than 400 nm.
- Only nanoparticles smaller than 500 nm can enter the circulatory system.

Scientists are discussing the relationship between particle sizes, metallic composition and their penetration into mesenteric lymphatic glands, but so far have reached no agreement [11]. In addition to being able to cross cell membranes and reach the blood and various organs because of their small size, nanoparticles have a bigger surface to volume ratio than larger particles. Therefore, more molecules of the metals are present on the surface, which may be one of the reasons why nanoparticles are generally more toxic than larger particles of the same composition [11].

The mechanisms of nanotoxicity are of different types: pulmonary and systemic inflammation, platelet activation, altered rate variability, and vasomotor dysfunction [29]. Nanotoxicity is reflected in the change in cell number in cytotoxicity tests reflects not just cell killing but also cell cycle arrest, which leads to a suppression of cell proliferation. Studies on cell cycle arrest induced by nanoparticles aid a better understanding of the reduction of viable cells.

**4.1 Particle surface**

Particle surface charge has a key-role on the cellular uptake of nanoparticles as well as how the particles interact with intracellular environment with organelles and biomolecules. Consequently, particle surface charge influences cytotoxicity. In vitro studies have shown that very small particles have more pathological and destructive power on the lungs rather than the same particles of smaller size due to their larger surface area, greater tendency to conjugate, and energy sustainability [10, 12, 30-33]. Two factors are related to cytotoxicity: (1) the atomic number of the element that increases; (2) the alteration of cell viability due to a function of particle surface charge, available binding site on a particle surface, and particle metal dissolution, but not of band-gap energy.

## 4.2 Molecular toxicity of metallic nanoparticles

Toxicity and other responses of metallic nanoparticles depend by the chemical composition, the prescribed dosage and substances used. Research has shown that a high dose of metals in small or big particles could be harmful to health, inducing also an increase of oxidative stress [30, 31, 38].

Chemical modification of the metallic particle surface has important effects on nanoparticles as they can react with metals. Iron can be affected by nanoparticles, which increases the induction of ROS in the free cell system. The surface modification of nanoparticles can reduce toxicity. Researchers have also shown that the toxicity of super paramagnetic iron oxide nanoparticles could be reduced by coating them with core-shell structures [30, 31, 38].

Most or all metallic particles produce free radicals in the free cell system and this ability causes oxidative stress, which gives rise to inflammation, cell destruction, and genotoxicity. The surface of metallic nanoparticles can activate the redox cycle and cause particle toxicity [38, 39].

## 4.3 Toxicity mechanisms of metallic nanoparticles in the respiratory tracts

The interaction between particle surface chemistry and the lung's surface-lining layer was investigated [34]. They found that, regardless of the nature of the surface, the metallic nanoparticles will be submersed into the lining layer after deposition in the small airways and alveoli. This displacement is promoted by the surfactant film itself as its surface tension falls temporarily to relatively low values [34, 35]. On the other hand, the reactive groups on a particle surface will certainly modify the biological effects. For instance, for silica it has been shown that surface modification of the quartz affects its cytotoxicity, inflammogenicity, and fibrogenicity. These differences are mainly due to particle surface characteristics [36]. The specific cytotoxicity of silica is strongly correlated with the appearance of surface radicals and reactive oxygen species (ROS), which is considered a key event in the development of fibrosis and lung cancer caused by this compound [37]. Although the type of particle does not seem to play an important role in whether it is embedded in the surfactant lining of the alveoli, the embedding process itself is crucial. Particle-cell interaction is possible only after the immersion of the particulates in the lining fluid, and research is needed to study this phenomenon in detail in relation to the inhaled nanoparticles. The reactive groups on silica nanoparticles influence their interaction with the lungs (or more generally with biological material) [37]. In some instances, it may be possible to predict the reactivity of the metallic nanoparticle surface. The scarcity of data, however, suggests that verifying these predictions by laboratory testing would be sensible. The degree of metallicity of a surface is the major feature used to estimate the toxicity. As well as size, it seems that the particle surface is critical in their absorption in the intestinal mucus [30, 31, 38].

## 5. Passage of nanomaterials through tissues

Metallic nanomaterials can enter the body via several portals, such as the skin, respiratory tract or gastrointestinal tract, or through parenteral administration [40, 41]. These organs protect the body from harmful environmental components. In other words, they are important organs in the transmission of nutrients, water, and oxygen. The skin acts as a barrier against the substances (apart from special elements such as oxygen for the retina and UV rays for vitamin D synthesis) [41-44]. Metallic nano-sized particles can enter and penetrate some organs such as the lungs, intestine, and skin. Some can penetrate the deepest layers of the skin (dermis). Their penetration depends on their size and nanoparticle surface features. It must be noted that in vitro tests must be carried out on nanoparticle toxicity before in vivo tests are performed. The digestive (gastrointestinal) tract has a close relationship with the environment. Metallic nanomaterials come into the body through the mouth and all nutrition is exchanged there, apart from gas. The histology of these three organs in relation to other places is different. The skin surface area of the body, which has an area of about 2 $m^2$ and a thickness of about 10 µm, is composed of keratin cells. These cells form a barrier against transmitting ions. The amount of penetration is related to organ, age, and other agents [41, 44]. The translocation of nanoparticles to the bloodstream allows interactions between nanomaterials, plasma and blood elements to take place. Nanoparticles' contact with plasma proteins leads to the formation of a "protein corona" [45], which may modify the biological and pharmacological properties of the metallic nanoparticles [46].

## 6. Toxicity of nanoparticles

Knowledge of the toxicity effects of these small substances is limited but is rapidly growing. Many studies have shown that some nanoparticles demonstrate toxicity in biological systems. Thus, research in the internal and external environment is needed; external studies can direct the internal studies. Some researchers have shown that most of the metallic nanoparticles can release active oxygen and cause oxidative stress and inflammation by the RES (reticoendothelial system). Acute toxicity resulting from nanoparticles has been investigated in vivo. The results indicate that toxicity depends on the size, coating, and chemical component of the metallic nanoparticles. Also, the systemic effects of nanoparticles have been shown in different organs and tissues. The effects on inflammatory and immunological systems may include oxidative stress or pre-inflammatory cytotoxin activity in the lungs, liver, heart, and brain. The effects on the circulatory system can include prethrombosis effects and paradox effects on heart function. Genotoxicity, carcinogenicity, and teratogenicity may occur as a result of the effects of metallic nanoparticles. Some nanoparticles could pass the blood-brain barrier and cause brain toxicity; of course, more studies are required [47, 48]. Due to the high loading of nanoparticles, macromolecule absorption will increase, so that they can cross through the digestive tract [49, 50].

Nanoparticles such as titania, zirconia, silver, diamonds, iron oxides, carbon nanotubes have been studied in diagnosis and treatment. Many of these nanoparticles may have toxic effects on cells. Most metal oxide nanoparticles may show toxic effects, but no toxic effects have been observed with biocompatible coatings. Biodegradable coating on metallic nanoparticles are also used in the efficient design of medical materials. Nanomaterials, even when made of inert elements like gold, become highly active at nanometer dimensions. In this regard nanotoxicology is fundamental to deal with the study and correct application of metallic nanomaterials. Nanotoxicity can lead to cell apoptosis or suppression of proliferation (via cell cycle arrest) because cells cannot overcome the stress and fix the damage. Different cell mechanisms may involve toxicity induced by exposure to metallic nanoparticles and nanomaterials. The alteration of cell signaling pathways induced by nanoparticles perturb cellular homeostasis leading to cellular injuries. When the mechanisms that induce which cell cycle phase could become arrested are multiple, the consequential suppression of proliferation reduces the cell number from one generation of cells to the next.

Expanding the knowledge of nanoparticle toxicity will facilitate designing of safer metallic nanocomposites and their application in a beneficial manner. Nanomaterials pose new challenges for biocompatibility due to the ability of nanomaterials to interact with the body on the cellular, subcellular and molecular levels.

**7. Nanomaterial toxicity in drug delivery systems**

Metallic nanoparticles can be used to transmit drug targets (as a drug or transmitter) or increase drug effectiveness [51]. Nanotechnology is also important, especially for detecting and treating cancer, although many problems have yet to be solved. Targeted gene delivery and transfection has been done both in vitro and in vivo with different types of metallic nanoparticles. In nanoparticle distribution in the body, immunological, pathological, pharmacological, and pharmacodynamic factors (time-base level of absorption, metabolism, and drug clearance) control the distribution of biological behaviors of nanoparticles in the body. In an in vivo environment, the fate of nanoparticles depends on these factors. After intravenous effusion of nanoparticles in vivo, nanoparticles are rapidly removed from the blood. This action is normally done by the immune and reticuloendothelial system without considering the particle features. Autoradiographic studies have shown that nanoparticles are usually concentrated in the liver and bone marrow.

Liver plays a key-role in the clearance of nanoparticles. The highest concentration of nanoparticles is found in liver cells. On the other hand, the concentration of these particles on mononuclear phagocytes causes the drug to keep away from the target cells. Several methods are available for preventing this event. One is magnetic directional guidance of intravenous particles out of the body.

One of the problems of using nanoparticles in pharmacology is their uptake by the mononuclear phagocytosis system as they exist in the liver and spleen. Besides the reduction in treatment effect, the uptake of nanoparticles in the liver may have a negative effect on liver function. The inflammatory responses by glycoprotein acid diffusion are caused by hepatocytes [52].

In pharmacology, the organs or cells are equally important, but nanoparticle effects on cells are important. For encapsulated drug activity, releasing of intracellular fluid is required; however, cell absorption for nanoparticles that are 20 nm or less in size do not need the endocytic mechanism. Chemical features such as surface load may affect intracellular nanoparticles. One of the effects of nanoparticle formulation is an increase in cell encapsulation. Use of metallic nanoparticles as drug transmitters may reduce combined drug toxicity. Usually, drug toxicity profiles are studied extensively while nanoparticle results are not described.

Another type of metallic nanoparticles are quantum dots, tiny semiconductor particles a few nanometres in size, having optical and electronic properties, made of binary compounds such as lead sulfide, lead selenide, cadmium selenide, cadmium sulfide, cadmium telluride, indium arsenide, and indium phosphide.

Quantum dots of 15-20 nm show that these particles migrate and concentrate in lymphatic nodes around the injection zone. Nanoparticles can be injected in veins in a colloid drug releasing system. They can also be injected in the muscle or used in oral or optical applications. With intravenous injections, quantum dots move to the reticuloendothelial system (RES), which releases them into the liver, spleen, brain, bone, and cardiac, renal, and respiratory systems. This distribution changes when the hydrophobicity and the surface load are corrected with the nanoparticle covers on different surfaces (same as surface correction with PEG and poloxamer). The nanoparticles are largely removed by mononuclear phagocytes after intravenous injection. On the other hand, it has been proved that the distribution model of nanoparticles in the body can affect its toxicity. In fact, changing the drug distribution and pharmacokinetics could help to develop a new model for drug effect and metabolism [52].

For example, doxorubicin cardiac and bone toxicity may be increased after combing this drug with smaller metallic nanoparticles. Accordingly, in a novel drug transmission system, the possibility of mononuclear phagocyte activation or inhibition should be considered. For example, with changing nanosuspension surface features, their effects can be improved even more. Nanoparticle adhesion is an important factor in bioapplication, drug absorption, and reduction of drug clearance [52, 53].

## 8. Nanomaterial toxicity mechanism in the blood brain barrier

One of the advantages of using nanoparticles in drug formulations is their potential for crossing the blood-brain barrier, although this function could have harmful effects. The nanoparticles have a toxic effect on cerebral endothelium cells. Of course, this is not true when applied to all nanoparticles of the same size. The physical features of biological materials and their ability to adhere to nanoparticles are important. When nanoparticles with different surface features are tested, neutral nanoparticles and anionic nanoparticles do not have any effect on the blood cerebral system, while a high density of cationic nanoparticles has toxic effects on the blood-brain barrier system. The surface load of metallic nanoparticles must be considered with regard to their toxicity effect and cerebral distribution profiles.

## 9. The potential role of nanoparticles' biological coating

A key role in the interaction of metallic nanoparticles with human physiology is played from "protein corona". Protein coronas act as the nanoparticles' "biomolecular fingerprints", influencing their pathophysiological effect [54] with their biochemical composition and structural conformation appearing to play key roles [55]. Interestingly, the corona is mainly composed of antibodies and complement proteins that normally facilitate the uptake of pathogens by phagocytes in a process much like that of opsonization [56]. On this basis, mapping the wear debris' corona could contribute to the engineering of better-tolerated materials.

In the last years, research on protein corona involve proteomics, immunology characterization and nanotoxicology, with many published articles exploring the effects of well-defined biological media (protein dispersions, sera, etc.) on well-characterized nanoparticles [57, 58]. Protein corona characterization of nanoparticles is based on single-step gradient centrifugations that are simple procedures based on the different densities of biological species and most exogenous nanoparticles. Upon their extraction, qualitative and quantitative assessments on nanoparticles, the examination of their biomolecular corona should provide a detailed characterization of the chemical-biological interaction of nanoparticles with the plasma proteins of human blood.

## 10. Nanotoxicology aspects of metallic nanoparticles on blood components

Different studies have pioneered and investigated the effects of metallic nanoparticles on vascular hemostasis and blood platelet function, as platelet-nanoparticle interactions are likely to play a major role in the hemocompatibility of metallic nanomaterials. These studies have shown that soluble and surface-bound carbon nanoparticles can stimulate platelet aggregation and increase thrombosis in vitro and in vivo [59, 60]. Metallic nanomaterials, such as silica nanoparticles [61], gold nanoparticles [62] and quantum dots [63], may also stimulate platelet activation. Interestingly, organic nanoparticles appear to be more platelet-compatible than inorganic nanomaterials [64], and this compatibility can also be increased by nanoparticle surface functionalization [62]. The molecular mechanisms involved in the interactions of platelets with nanomaterials are multifactorial, but they may be triggered by the generation of reactive oxygen and nitrogen species such as peroxynitrite [61, 65], and depend on the modification of receptors, transport systems and intraplatelet transduction mechanisms [66]. Unique test systems have been designed for studying the effects of metallic nanoparticles on platelets. The use of a quartz crystal microbalance facilitates highly sensitive measurement of platelet microaggregate formation during interactions among nanoparticles, platelets and other cells under flow conditions [59, 62, 63, 66, 67]. An understanding of the subcellular mechanisms involved in metallic nanomaterial hemocompatibility, as well as the availability of sensitive methods for studying such interactions, will be of great assistance when developing diagnostic and therapeutic options using multifunctional metallic nanomaterials.

## 11. Nanomaterials and their commercialization

It is also important to point out that there are compelling differences between various medically relevant metallic nanomaterials. Therefore, the R&D as well as the registration of novel metallic nanodiagnostic/nanotherapeutic agents with the FDA and European Medicines Agency is likely to proceed on a case-to-case basis. It is worth commenting on risk and safety assessment associated with the use of metallic nanomaterials. Both the FDA and the EU have already set up robust schemes for the approval and legislation of new metallic nanopharmaceuticals; these schemes appear to be working well and provide comprehensive pharmacological/toxicological profiles of novel metallic nanoparticles. A challenge that requires very careful consideration is the environmental presence and persistence of engineered metallic nanoparticles. It is likely that organic, biodegradable nanoparticles will be of lesser concern, as they will be degraded by metabolic pathways [68]. In contrast, inorganic, non-biodegradable nanoparticles, including multifunctional magnetic nanoparticles, may persist for considerable periods and result in prolonged exposure of humans, animals and the environment with still-to-be-determined consequences. This could potentially add to the pool of engineered nanomaterials present in the environment, which is continuously delivered by nanoproducts produced by the food, agriculture and cosmetics industries [68, 69, 70, 71]. The outcome of biological interactions between metallic nanomaterials and xenobiotics present in the environment may also be of concern [72]. Therefore, serious assessment of the environmental risks associated with the growing presence of nanotechnology metal composite products in the environment most definitely needs to be carefully considered [73, 74, 75].

## 12. Conclusions

Nanotoxicology emerged to better understand the impact of nanoparticles and nanomaterials on environmental and human health and help us move toward making safer and more effective materials. In vitro studies are fundamental to identify biochemical and molecular mechanisms of cytotoxicity and to investigate the complexities of toxicokinetics and toxicodynamics of nanoparticles' biodistribution.

Metallic nanomaterials, which are used in the field of medical science, have been discussed and their toxicity effects investigated. It is obvious that most metallic nanoparticles considered in this review are toxic and can influence the body's cells. The biocompatible coatings improve the performance of these nanoparticles, reduce their toxicity, and do not result in negative effects on cells. The knowledge of processes, including absorption, distribution, metabolism and excretion, as well as careful toxicological assessment is critical in order to determine the effects of metallic nanomaterials in humans and other biological systems. In vitro studies provide insights to hazard identification of nanomaterials which can induce to further studies on animal subjects. They are also the first step in identifying occupational risk assessment. Expanding the knowledge of nanoparticle toxicity will facilitate designing of safer nanocomposites and their application in a beneficial manner. To achieve this aim, research on metallic nanotoxicity must become more generalizable. Given the huge number of potential interaction pairings between industrially relevant metallic nanomaterials and biological systems, a case-by-case approach to assessing nanotoxicity is unlikely to keep pace with advances in nanotechnology. Two approaches already beginning to be adopted by the field can improve the generalizability of nanotoxicity research. The first approach seeks to identify fundamental mechanisms of metallic nanomaterial-biological interaction and toxicity by probing the molecular character of the metallic nano-bio interface directly. This approach requires the continued development and application of highly sensitive and selective analytical tools, capable of probing the chemically complex environments of many nanotoxicity studies. The second approach seeks to identify generalizable relationships between metallic nanomaterial structure and biological activity by applying high-throughput analyses paired with mathematical and computational approaches to analyze toxicological data sets.

**Acknowledgements**

This article is based upon work from COST Action CA 17140 "Cancer Nanomedicine from the Bench to the Bedside" supported by COST (European Cooperation in Science and Technology).